\documentstyle[preprint,epsfig,aps]{revtex}
\begin{document}
\preprint{SNUTP 99-028}
\title{Boundary Effects on Dynamic Behavior of
 Josephson-Junction Arrays}
\author{M.Y. Choi, Gun Sang Jeon, and Mina Yoon~\cite{MY}}
\address{Department of Physics and Center for Theoretical Physics\\
Seoul National University\\
Seoul 151-742, Korea}
\maketitle
\draft

\begin{abstract}
The boundary effects on the current-voltage characteristics
in two-dimensional arrays of resistively shunted Josephson junctions 
are examined. 
In particular, we consider both the conventional boundary 
conditions (CBC) and the fluctuating twist boundary conditions (FTBC), 
and make comparison of the obtained results.
It is observed that the CBC, which have been widely adopted in 
existing simulations,
may give a problem in scaling, arising from rather large boundary effects;
the FTBC in general turn out to be effective in reducing the finite-size
effects, yielding results with good scaling behavior.
To resolve the discrepancy between the two boundary conditions, 
we propose that the proper scaling in the CBC should be performed 
with the boundary data discarded: This is shown
to give results which indeed scale well and are the same as those 
from the FTBC.

\end{abstract}
\bigskip
\thispagestyle{empty}

%{\tiny \noindent 
%74.50.+r Proximity effects, weak links, tunneling phenomena, and Josephson effects\\[-\baselineskip]
%74.25.Fy Transport properties (electric and thermal conductivity, thermoelectric effects, etc.)\\[-\baselineskip]
%74.40.+k Fluctuations (noise, chaos, nonequilibrium superconductivity, localization, etc.)\\[-\baselineskip]
%64.60.Ht Dynamic critical phenomena }

%\begin{multicols}{2}
%\narrowtext

%\section{Introduction}
During the past decades superconducting arrays have been extensively
studied both theoretically and experimentally~\cite{proceeding}.
The superconducting arrays in equilibrium are well described by 
the $XY$ model, which exhibits the 
Berezinskii-Kosterlitz-Thouless (BKT) transition 
at a nonzero temperature in two dimensions~\cite{BKT}.
Theoretically this transition is manifested by the characteristic behavior of static
quantities such as the universal jump of the helicity modulus. 
On the other hand, the standard experimental probe of the transition
is to measure the dynamical response to the external current
driving~\cite{Exp}.
The need for the analysis of the experimental data
together with the interest in the dynamic properties themselves
has motivated extensive theoretical examinations of the dynamics
of the system, where numerical simulations 
have been of great help.

It has been recognized that the 
boundary conditions employed in those simulations, performed in finite systems,
may affect the results in a crucial way, which makes it necessary to
examine their effects.
Most existing dynamic simulations of the system in the absence of driving currents have been performed 
with periodic boundary conditions along both directions;
under driving currents, the boundary conditions along the current-injected direction are changed
to be free
whereas periodic boundary conditions are retained along the perpendicular direction.
Such conventional boundary conditions (CBC), however, are known to produce rather large
edge effects, and the system is expected to approach the
thermodynamic limit rather slowly,
compared with that in the busbar boundary conditions~\cite{Simkin}.
Of much interest here are the recent results on the dynamical exponent.
Under the periodic boundary conditions, remarkably different values of the dynamical exponent 
have been obtained in the resistively shunted Josephson junction (RSJ) dynamics and 
in the time-dependent Ginzburg-Landau (TDGL) dynamics~\cite{Jose,Kim}, suggesting 
that the two kinds of dynamics may be inherently different.
In particular the dynamical exponent in the RSJ dynamics turns out to be
much smaller than expected and apparently inapplicable to the flux-noise
experiment~\cite{Shaw}. 
On the other hand, no difference has been observed 
when the fluctuating twist boundary conditions (FTBC), 
forcing the vortex interaction to be periodic~\cite{Ol}, is employed~\cite{Kim};
this implies that the difference in the dynamical exponent stems from the 
employed boundary conditions rather than from the intrinsic property of
the dynamics.

This paper is to clarify the boundary effects and thus to resolve the discrepancy
between the two types of the boundary conditions.
For this purpose, we investigate in detail the current-voltage $(IV)$
characteristics and the corresponding dynamic critical behaviors 
of the system in the CBC and in the FTBC.
The finite-size-scaling analysis is performed to reveal that the
two types of the boundary conditions yield different values of the dynamical exponent.
In particular, the data obtained in the CBC are found 
not to display proper scaling behavior.
We show that the discrepancy has its origin in the rather large 
boundary effects in the CBC
and give a simple proposal that the proper scaling analysis 
in the CBC should be performed with the boundary data discarded.
It turns out that the resulting $IV$ characteristics indeed exhibit
good scaling behavior and are consistent with those from the FTBC,
confirming that the poor scaling behavior originates from the peculiar
boundary effects in the CBC and that correct description of the
dynamic behavior can be achieved by appropriate extraction of the
bulk properties from the data.

%This paper is organized as follows:
%In Sec.~\ref{set:IV}, we compute the $IV$ characteristics 
%of the superconducting arrays under the two boundary conditions, 
%CBC and FTBC, and perform the finite-size scaling analysis
%of the $IV$ data. 
%It is demonstrated that the data from the CBC do not collapse within 
%the one-parameter scaling curve while those from the FTBC display
%nice scaling behavior.
%In Sec.~\ref{set:BE} the failure of the scaling is attributed to
%the large boundary effects produced in the CBC.  Indeed good
%scaling behavior is shown to be restored when the boundary data are discarded
%in the analysis, thus confirming the crucial role of boundary conditions
%in the finite-size scaling analysis.
%Finally, Sec.~\ref{set:CON} summarizes the main results of this
%paper.
%==================================================

%\section{Finite-Size Scaling Analysis of Current-Voltage Characteristics}
%\label{set:IV}
%==================================================
%In this section, the $IV$ characteristics of an
%$N{\times}N$ RSJ array are computed.
%We consider two different boundary conditions:
%the CBC, consisting of free boundary conditions 
%along the current-injected direction and 
%periodic boundary conditions along the perpendicular direction,
%and the FTBC, where new variables $\Delta_x$ and $\Delta_y$ 
%are introduced with the periodic
%boundary conditions imposed in both directions.

We begin with the set of equations of motion describing
the dynamics of an RSJ array with
critical current $I_c$ and shunt resistance $R_s$:
\begin{equation}
\label{eq:dyn1}
{\sum_j}' \left[ {\hbar \over 2eR_s} {d \phi_{ij} \over dt}
 + I_c \sin\phi_{ij} + L_{ij}\right]
= I_i^{\rm ext} .
\end{equation}
Here $\phi_{ij}\equiv \phi_i - \phi_j$ is the phase difference
across a junction with
$\phi_i$ being the phase of the superconducting order parameter 
on grain $i$ (at position ${\bf r}_i$),  
$I_i^{\rm ext}$ is the external current fed into grain $i$,
and the primed summation runs over the nearest neighbors of grain $i$.
The thermal noise current $L_{ij}$ is assumed to be a white noise
satisfying
\begin{equation}
\langle L_{ij}(t+\tau) L_{kl} (t) \rangle
= {2 k_B T \over R_s} \delta (\tau) (\delta_{ik}\delta_{jl} -
\delta_{il}\delta_{jk})
\end{equation}
with the angular bracket denoting an ensemble average and $T$
being the temperature.
In the CBC, uniform current $I$ is fed into each grain at 
one edge ($x=0$) and extracted from each grain at the other edge ($x=N$), 
leading to $I_i^{\rm ext} = I (\delta_{x_i0} - \delta_{x_iN})$~\cite{Choi}.
The boundary conditions for the phase variables are taken to be 
free in the $x$ direction and periodic in the $y$ direction. 
The average voltage across each junction is given by the relation
\begin{equation} \label{eq:V1}
V = {\hbar \over 2eN^2} \sum_{y=1}^N \left(
\left\langle {d \phi_{(0,y)} \over dt} \right\rangle_t -
\left\langle {d\phi_{(N,y)} \over dt} \right\rangle_t \right),
\end{equation}
where $\langle\ldots\rangle_t$ stands for the time average.
In the FTBC, on the other hand, 
periodic boundary conditions are imposed on $\{\phi_i\}$ 
in both directions, yielding
the equations of motion in the form~\cite{Kim}
\begin{equation}  \label{eq:dyn2}
{\sum_j}' \Big[ {\hbar \over 2eR_s}
{d \phi_{ij} \over dt} 
 + I_c \sin (\phi_{ij} - {\bf r}_{ij} \cdot {\bf \Delta}) 
+ L_{ij}\Big]=0
\end{equation}
with ${\bf r}_{ij} \equiv {\bf r}_i - {\bf r}_j$.
The dynamics of the twist variables 
${\bf \Delta} \equiv (\Delta_x,\Delta_y)$ 
is governed by 
\begin{eqnarray}
%\label{eq:dyn3}
\frac{\hbar}{2eR_s}\frac{d\Delta_{x}}{dt} &=& \frac{I_c}{N^2}
\sum_{\langle ij \rangle_{x}}
\sin(\phi_{ij} -\Delta_{x}) - {I \over N} + L_{\Delta_{x}} , \nonumber \\ 
\label{eq:dyn4}
\frac{\hbar}{2eR_s}\frac{d\Delta_{y}}{dt} &=& \frac{I_c}{N^2}
\sum_{\langle ij \rangle_{y}}
\sin(\phi_{ij} -\Delta_{y})  + L_{\Delta_{y}} ,
\end{eqnarray}
where $\sum_{\langle ij \rangle_{x(y)}}$ denotes the summation over 
all nearest-neighboring pairs in the $x(y)$ direction and
the thermal noise current $L_{\Delta_{x(y)}}$ satisfies
\begin{equation}
\langle L_{\Delta_{x(y)}}(t+\tau) L_{\Delta_{x(y)}} (t) \rangle =
 {2 k_B T \over N^2 R_s} \delta (\tau). 
\end{equation}
The average voltage can then be obtained from the time evolution of the twist
variable $\Delta_x$:
\begin{equation}
V = - {\hbar \over 2e} 
\left\langle {d \Delta_x \over dt} \right\rangle_t.
\end{equation}

To obtain the $IV$ characteristics in the two boundary conditions, we
integrate directly the coupled equations of motion given by Eq.~(\ref{eq:dyn1})
for the CBC and Eqs.~(\ref{eq:dyn2}) and (\ref{eq:dyn4}) for the FTBC, with the time
step $\Delta t = 0.05$ (in units of $\hbar /2 e R_s I_c$).
The data are averaged over 30 independent runs while
in each run the averaging time is chosen in such a way that at least 25 
vortices are created. 
We 
%consider several system sizes ranging from $N=4$ to $N=64$ 
%to clarify size effects, and 
concentrate on the temperature $T=0.84$ (in units of
$\hbar I_c / 2e$), which is below the BKT transition temperature $T_c
\approx 0.89$~\cite{Ol2}, and
henceforth write the current and the voltage in units of 
$I_c$ and $I_c R_s$, respectively. 

Figure~\ref{fig:IV} presents the resulting $IV$ characteristics 
in the CBC and in the FTBC.
We have considered several system sizes ranging from $N=4$ to $N=64$,
to clarify size effects.
At high currents the average voltage drop
is found to be almost independent of the size in both boundary conditions.
As the current is decreased, however, the system undergoes a crossover from the
nonlinear-resistance regime to the linear-resistance regime, exhibiting
size-dependence.
The crossover current apparently decreases with the increase of the size, suggesting that
the linear resistance at low currents reflects the finite-size
effects~\cite{Simkin,Wallin} rather than the lattice effects~\cite{Choi}.
Here one can observe two striking differences between the
$IV$ characteristics in the CBC and those in the FTBC:
First, at low currents the voltage drop in
the CBC is much higher than that in the FTBC.
Furthermore, the ratio of the former to the latter grows larger as
the system size is increased.
Secondly, the decrease of the crossover current with the system size is
faster in the FTBC than in the CBC; this implies that the FTBC is more effective in
reducing the finite-size effects.  
%
%The resulting $IV$ characteristics are not shown here because
%similar figures can be found in the existing studies~\cite{Simkin,Kim,Choi}.

%==================================================
%\section{Finite-Size scaling}
%\label{set:FSS}
We now analyze the $IV$ data  by means of the finite-size scaling
method, and investigate the dynamic critical behavior both in the CBC and
in the FTBC.
We assume the dynamic scaling form~\cite{Kim,Wallin}
\begin{equation}
N R^{1/z} = f(N I),
\label{eq:FSS}
\end{equation}
where $R$ is the resistance of the system and $z$ is the dynamical exponent.
The dimensionless scaling function $f(x)$ is expected to possess the asymptotic behavior:
$f(x) \propto {\rm constant}$ for small $x$ and $f(x) \propto x$ for large $x$.
%\begin{equation} \label{eq:asym}
%f(x) \propto \left\{
%\begin{array} {cl}
%\hbox{const.}& \hbox{        for small $x$},\\
%x & \hbox{        for large $x$}.
%\end{array}
%\right.
%\end{equation}
The small-$x$ behavior follows from the behavior of the linear resistance for $I \rightarrow 0$:
\begin{equation}
R_L \sim N^{-z}
\label{RL}
\end{equation}
whereas the finiteness of the resistance in the limit $N\rightarrow \infty$ 
elucidates the behavior of $f(x)$ for large $x$.

We first fit the linear resistance $R_L$ in the CBC to the form 
in Eq.~(\ref{RL}), and display in the inset of Fig.~\ref{fig:FSS}(a) 
the log-log plot of the linear resistance versus the system size, 
which shows that the linear resistance is well described by Eq.~(\ref{RL})
with the dynamical exponent $z=0.91 \pm 0.07$. 
Note, however, that this is presumably deceptive since
the value of $z$ less than two is likely to be erroneous,
especially, below the transition temperature:
In a two-dimensional Josephson-junction array, the value is
generally believed to be $z=2$ at the transition temperature 
and to grow larger as the temperature is lowered~\cite{Wallin}.
Indeed the scaling plot with the value $z=0.91$, which is presented 
in Fig.~\ref{fig:FSS}(a), manifests that the system in the CBC does not 
follow the one-parameter dynamic scaling given by Eq.~(\ref{eq:FSS}).

We next examine the $IV$ data in the FTBC,
and fit the corresponding linear resistance $R_L$.
This yields the dynamical exponent $z=2.73 \pm 0.02$, 
as shown in the inset of Fig.~\ref{fig:FSS}(b),
which is quite different from the value in the CBC 
and indeed consistent with the results of previous studies~\cite{Kim,Wallin}.
In particular, unlike the result of the CBC,
the scaling plot shown in Fig.~\ref{fig:FSS}(b) reveals that
the scaling form in Eq.~(\ref{eq:FSS}) is observed well
in the FTBC.  Namely, the $IV$ data in the
nonlinear-resistance regime as well as in the linear-resistance regime
collapse nicely into a single curve and the scaling function follows the expected
asymptotic behaviors.

%In the next section we will investigate in detail 
%the origin of the discrepancy and propose its possible resolution.

%==================================================
%\section{Boundary Effects}
%\label{set:BE}
%==================================================
Both the values of the dynamical exponent and the resulting scaling plots
indicate that the CBC are not appropriate for describing 
the dynamical behavior of the system whereas the FTBC give a correct
description.
This discrepancy is likely to have its origin in the
effects of the boundary, which is formed for the purpose of the 
current injection in the CBC but is absent in the FTBC.
To find a clue to the boundary effects,
we examine the distribution of the voltage drop along the $x$
direction, which is, according to
Eq.~(\ref{eq:V1}), proportional to the gradient of the phase profile.
The phase profile 
\begin{equation} \label{eq:profile}
\Delta \phi(x) \equiv 
{1 \over N} \sum_y [\phi_{(x,y)} (\tau)-\phi_{(x,y)}(0)],
\end{equation}
where $\tau$ is the average time, 
has been obtained for several system sizes and currents,
%is displayed in Fig.~\ref{fig:profile} for the system of size $N=32$ at current $I=0.04$.
%Figure~\ref{fig:profile} 
which turns out to be essentially the same as the profile
obtained in Ref.~\cite{Simkin}.  
It demonstrates that the voltage drop across a few
columns near the boundary is comparable to, or even larger than, that
across the rest of the system. 
Such tendency becomes prominent at lower currents,
and leads to an overestimation of the voltage drop in the system,
especially at low currents; this may 
have significant effects on the dynamic behavior of a system
of finite size~\cite{Simkin}.

To make clear that the boundary effects are the main origin
of the failure of the dynamic finite-size scaling in the CBC,
we consider the $(N{+}10)\times N$ array, consisting of 
$N{+}10$ columns and $N$ rows under the CBC,
and calculate the voltage drop between the grains at
$x=5$ and $x=N{+}5$.
Namely, we propose that the proper scaling in the CBC should be performed 
with the boundary data discarded,
which helps to estimate correctly the voltage drop in the bulk~\cite{comment}.
Here the number five has been chosen since the phase profile reveals
that the boundary effects are severe, i.e., the anomalous voltage drop 
is substantial, across the five columns at each edge,
apparently regardless of the system size.
The $IV$ characteristics for several sizes, obtained in this scheme,
are plotted in Fig.~\ref{fig:IVMCBC},
which shows marked difference from Fig.~\ref{fig:IV}(a):
%that computed in the original CBC.
The voltage drop is significantly reduced at small currents, 
resulting in the $IV$ characteristics 
which look almost the same as those obtained from the FTBC.

Indeed the finite-size scaling analysis, shown in the inset of Fig.~\ref{fig:FSSMCBC},
leads to the dynamical exponent $z=2.84 \pm 0.06$, which is consistent with 
that obtained from the FTBC.
Further, with this value of the dynamical exponent, 
the $IV$ data in Fig.~\ref{fig:IVMCBC}
collapse remarkably well into a single curve, 
as illustrated in Fig.~\ref{fig:FSSMCBC}.
>From these results we thus reach the conclusion that the 
CBC give a correct
description of the dynamical behavior of the Josephson-junction
array only when the rather large voltage drop 
near the boundary is discarded in the analysis. 
Such boundary effects should be irrelevant in the thermodynamic limit, but
can affect significantly the properties of a finite system,
generating substantial errors in the scaling analysis.
%
%It is of interest to note that under periodic boundary conditions 
%the values of $z$ far smaller than expected were also
%obtained from the scaling analysis of the relaxation time
%in the RSJ model both above the BKT transition temperature $T_c$~\cite{Jose}
%and below $T_c$~\cite{Kim}.
%In contrast the same analysis of the TDGL model was found to
%give $z \approx 2$ at the same temperature~\cite{Jose};
%this led to the conclusion that the TDGL model describes the flux-noise 
%experiment~\cite{Shaw} better than the RSJ model.
%In a subsequent study of the low-temperature regime ($T\lesssim T_c$) 
%via the FTBC, however, 
%both models were shown to possess the same value of 
%$z\, (\gtrsim 2)$~\cite{Kim}.
%Accordingly, the erroneous value of $z$ obtained in the RSJ model
%was proposed to be an artifact of the periodic boundary conditions 
%rather than the inherent property of the RSJ model.
Taking into account that the periodic boundary conditions are naturally
generalized to the CBC in the presence of driving currents, we can infer
that the erroneous value of $z$ obtained 
in Ref.~\cite{Jose} also has its origin in the 
boundary conditions rather than in the RSJ dynamics itself.

%==================================================
%\section{Summary}
%\label{set:CON}
%==================================================
In summary, we have studied the boundary effects on the 
dynamical behavior of a two-dimensional RSJ array, both
in the CBC and in the FTBC.
%In particular, we have investigated in detail the current-voltage 
%characteristics and the corresponding dynamic critical behaviors 
%of the system in the conventional boundary conditions and 
%in the fluctuating twist boundary conditions.
The finite-size-scaling analysis has been performed to reveal 
%that the
%two types of the boundary conditions yield different 
not only the discrepancy in the value of the dynamical exponent 
but also the lack of scaling in the CBC,
%obtained in the the conventional boundary conditions
%are found not to exhibit proper scaling behavior
%while the fluctuating twist boundary conditions
%yielded results displaying good scaling behavior.
%It has been demonstrated that the discrepancy arises 
steming from the rather large boundary effects.
%in the conventional boundary conditions.
We have thus given a simple proposal that in the CBC
the proper scaling should be performed with the boundary data discarded,
and shown that such removal of the boundary effects indeed restores 
the correct scaling behavior.
%in the CBC~\cite{comment}
This confirms the crucial role of boundary conditions
in the finite-size scaling analysis and
resolves the discrepancy between the two types of the boundary conditions.

%\acknowledgements
We thank B.J. Kim for useful discussions and acknowledge the partial support
from the Seoul National University Research Fund and from the 
Ministry of Education through the BK21 Program.
G.S.J. was also supported by the Korea Science and Engineering 
Foundation postdoctoral fellowship.

%===========================================================================

\begin{center}

\begin{figure}
\vspace{0.5cm}
\centerline{\epsfig{width=10cm,file=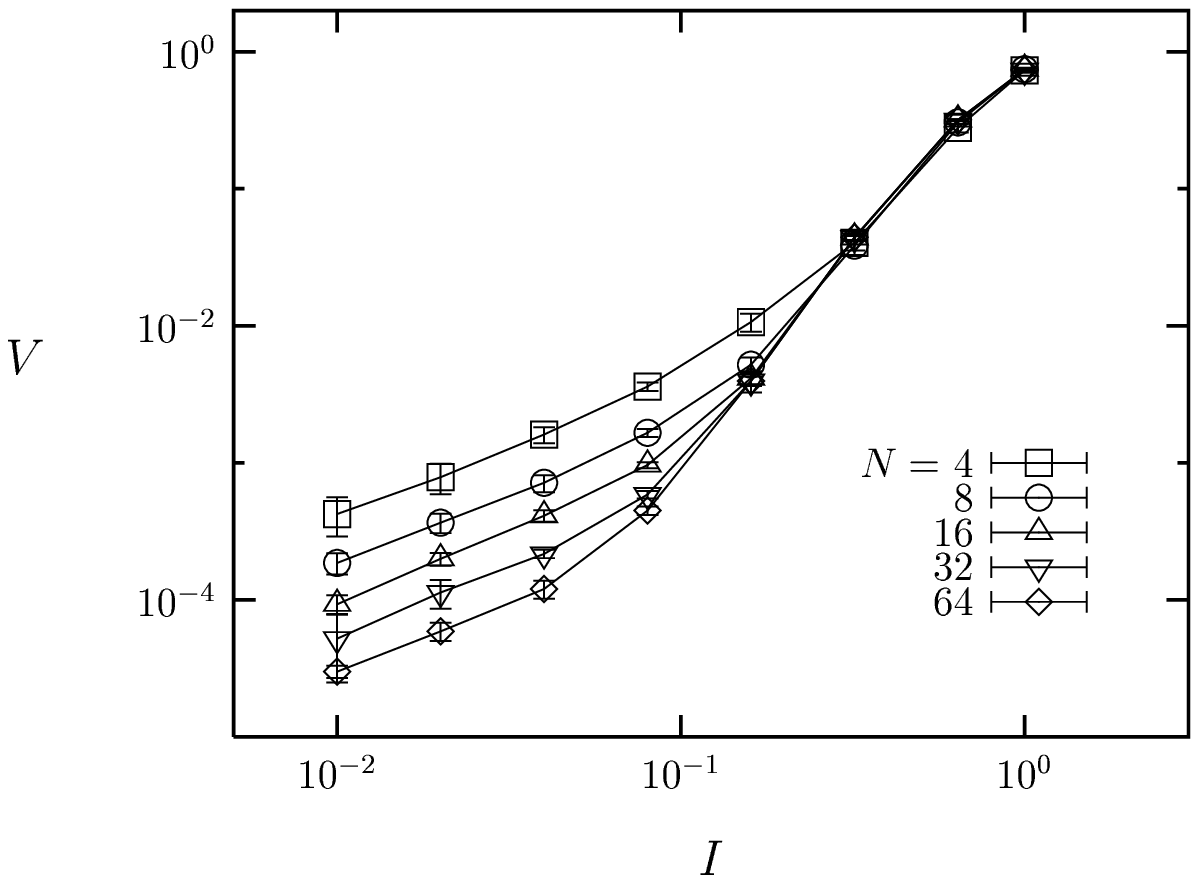}}
\centerline{\large (a)}
\vspace{1cm}
\centerline{\epsfig{width=10cm,file=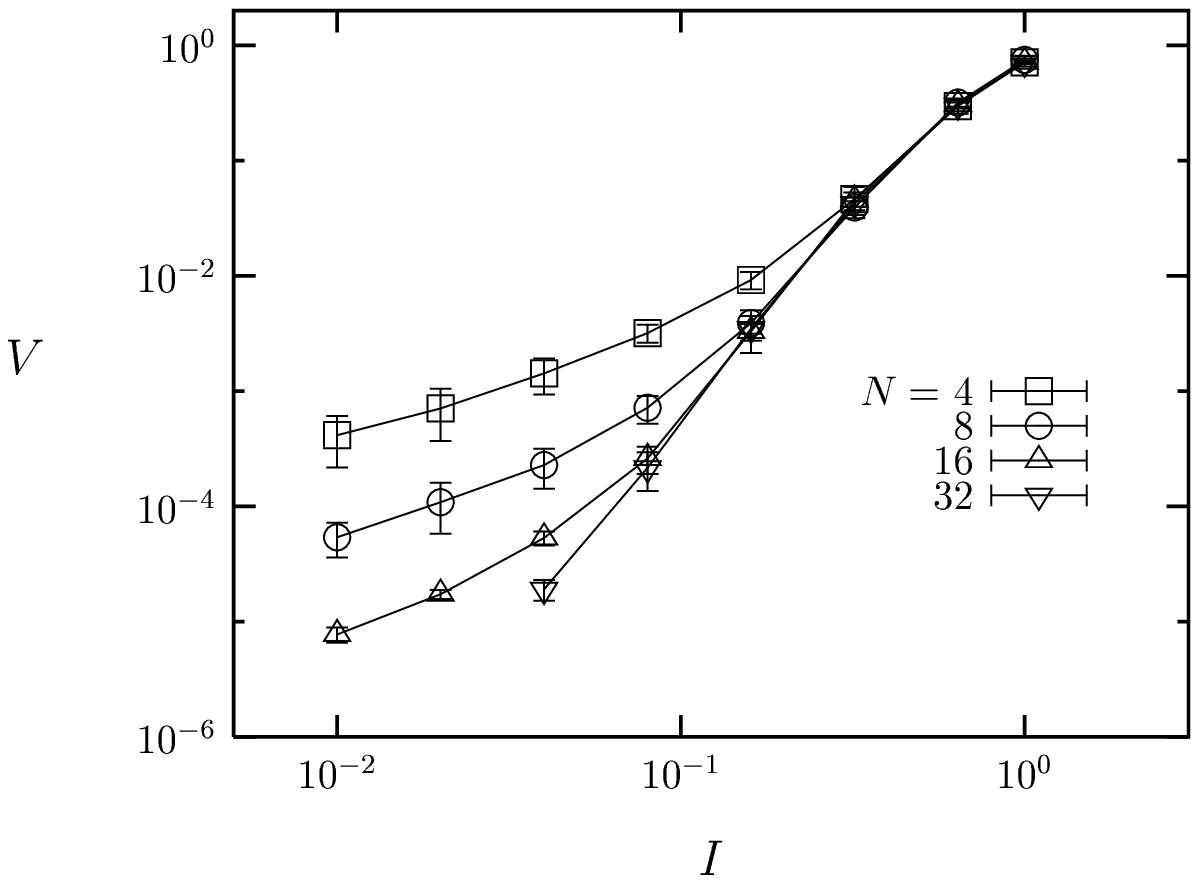}}
\vspace{-0.3cm}
\centerline{\large (b)}
\vspace{0.5cm}
\caption
{$IV$ characteristics of $N{\times}N$ square Josephson-junction arrays
at $T =0.84$ for several sizes (a) in the CBC and (b) in the FTBC.
Lines are merely guides to the eye.}
\label{fig:IV}
\end{figure}

\newpage
\begin{figure}
\vspace*{0.5cm}
\centerline{\epsfig{width=12cm,file=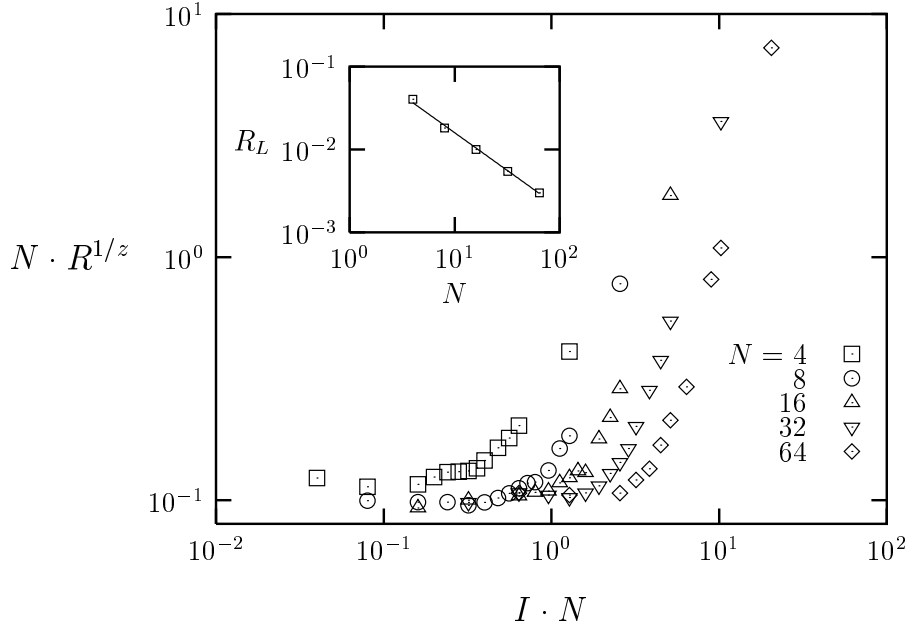}}
\centerline{\large (a)}
\vspace*{0.5cm}
\centerline{\epsfig{width=12cm,file=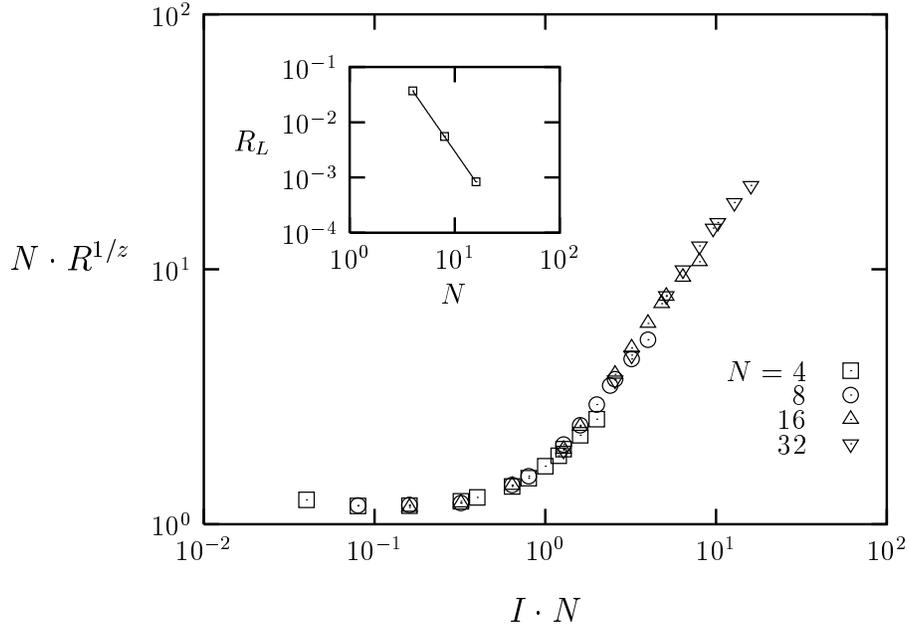}}
%\vspace{-0.3cm}
\centerline{\large (b)}
\vspace{0.5cm}
\caption
{Scaling plots of the $IV$ data 
(a) in the CBC with $z=0.91$ and (b) in the FTBC with $z=2.73$. 
Each inset displays the log-log plot of the linear resistance versus
the size and its power-law fitting curve, from which the value of the dynamical
exponent $z$ is estimated.
}
\label{fig:FSS}
\end{figure}

\newpage
\begin{figure}
\centerline{\epsfig{width=10cm,file=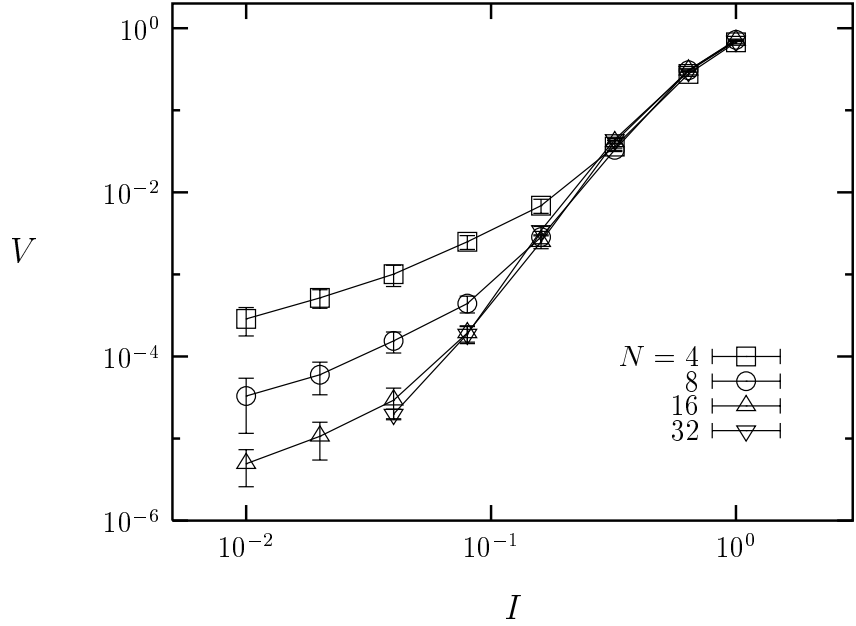}}
\vspace{0.5cm}
\caption
{$IV$ characteristics of $(N{+}10){\times}N$ square Josephson-junction arrays
at $T =0.84$ for several sizes in the CBC. 
The voltage is measured between the grains at $x=5$ and at $x=N{+}5$. 
Lines are merely guides to the eye.}
\label{fig:IVMCBC}
\end{figure}

\begin{figure}
\centerline{\epsfig{width=12cm,file=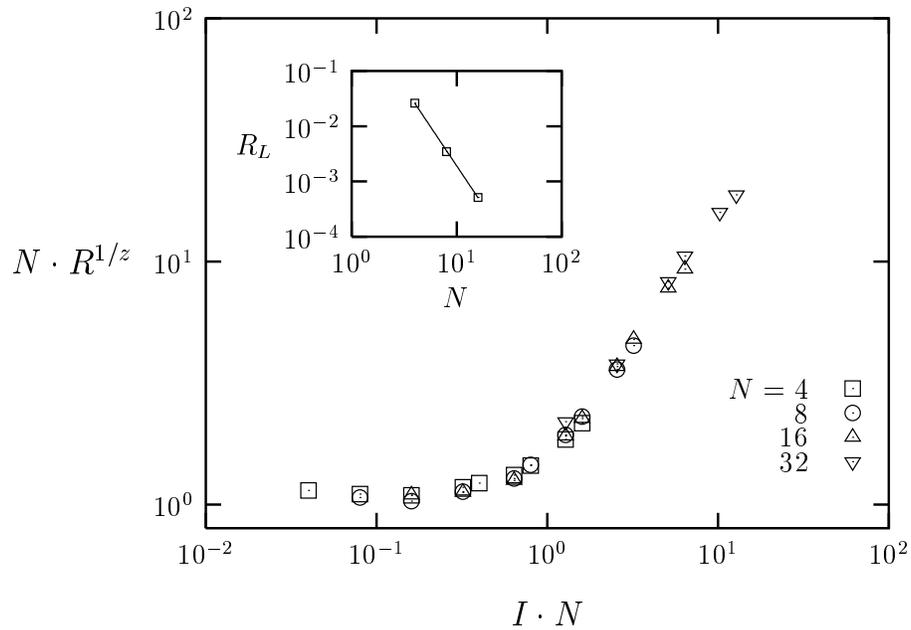}}
\vspace{0.5cm}
\caption
{Scaling plot of the $IV$ data presented in Fig.~\ref{fig:IVMCBC},
with $z=2.84$. 
Inset: the log-log plot of the linear resistance versus
the size and its power-law fitting curve.
}
\label{fig:FSSMCBC}
\end{figure}

%\begin{figure}
%\vspace{1cm}
%\centerline{\epsfig{width=6cm,file=fig1.eps}}
%%
%%\vspace{1cm}
%
%\caption
%{Schematic diagram of an $N{\times}N$ square Josephson-junction array 
%(corresponding to the case with $N=4$) in the CBC,
%which consist of the periodic and free boundary conditions 
%for phases in the directions perpendicular and parallel to
%the current flow, respectively.}
%\label{fig:array}
%\end{figure}

%\newpage

%\vspace{1cm}

%\begin{figure} 
%\centerline{\epsfig{width=11mc,file=fig4.eps}}
%\vspace{0.5cm}
%\caption
%{Phase profile, which is defined by Eq.~(\ref{eq:profile}), 
%as a function of the position along the direction of the current
%in the 32{$\times$}32 array, driven by the current $I = 0.04$ 
%under the CBC. 
%}
%\label{fig:profile}
%\end{figure}
%

%\vspace{1cm}

\end{center}
%\end{multicols}


\begin{references}
\bibitem{comment}
%Whereas the currents fed or extracted make the grains at the edges 
%$x=0$ and $x=N$ different from those inside, no such distinction is 
%present along the $y$ direction.  Accordingly, 
Here the boundary conditions 
along the transverse direction are not expected to have significant 
effects on the voltage drop along the current-injected direction
in Eq. (3), which is averaged over the transverse direction.  
Thus the periodic boundary conditions 
may be maintained along the transverse direction, without affecting 
qualitatively the scaling behavior, as manifesteded by Fig.~\ref{fig:FSSMCBC}.
\end{references}
\end{document}